# Quasi-cyclic self-dual codes of length 70

Alexandre Zhdanov
Voronezh, Russia
a-zhdan@vmail.ru

***Abstract*— In this paper we obtain a number of [70,35,12] singly even self-dual codes as a quasi-cyclic codes with m=2 (tailbitting convolutional codes). One of them is the first known code with parameters Beta=140 Gamma=0. All codes are not pure double circulant i.e. could not be represented in systematic form.**

***Keywords—convolutional encoding, quasi-cyclic codes weight enumeratorg, double circulant***

## I. INTRODUCTION

A linear binary code $C(n,k,d)$ is a subspace of $F_2^n$ of dimension $k$. The $F_2$ is a field of two elements: 1,0, where the summation is a logical XOR and multiplication is a logical AND. The codeword weight $d$ is a minimal number of non-zero component in any codeword of code $C$. The quasi-cyclic code is a code for which every cyclic shift of a codeword by $m$ symbols yields another valid codeword, where $m>1$. The quasi-cyclic code of $R=1/m$ consists of $m$ circulants. A circulant is a square matrix where the next row is obtained by one element cyclically shifting to the right the previous row. The cyclically shifting to the left will result an inverse circulant. The tailbitting convolutional code of $R=1/2$ is a quasi-cyclic code with $m=2$, where the columns of the circulants mixed to form a compact mixed polynomial string. The mixed polynomial string is a non-zero part of the generator matrix row. Self dual codes are a powerful class of codes. Self-dual code $C$ is a code with coding rate $R=1/2$, where the inner product of any two rows in a generator matrix $G$ gives 0. In other words: $C=C^{\perp}$, where $C^{\perp}$ is a dual code. All codeword's of binary self-dual code has even weight. If all codewords weights $\equiv 0(\text{mod}\,4)$ the code is called doubly even, if all codewords weights $\equiv 2(\text{mod}\,4)$ the code is called singly even. The code is called extremal if the minimum weight of the codeword meets the following bond: $d \le 4\lfloor n/24 \rfloor + 6$ if $n \equiv 22(\text{mod}\,24)$ and $d \le 4\lfloor n/24 \rfloor + 4$ otherwise [5, 6]. We refer the reader to [7] for details.

Let us consider the convolutional codes and its taps are described by the polynomials (Type $A_0$ [1]). In this case the generator matrix is obtained for example by cyclically shift of the mixed polynomial string $(p_0,q_0,p_1,q_1,\cdots,p_{K-1},q_{K-1},0,0,\cdots 0)$ with step 2 or use another form of generator matrix $G=[P\,|\,Q]$, , where $P$ and $Q$ are circulants $k \times k$ with top row $(p_0,p_1,\cdots,p_{K-1},0,0,\cdots 0)$ and $(q_0,q_1,\cdots,q_{K-1},0,0,\cdots 0)$ respectively. When one circulant is an identity matrix $I$ the construction $G=[I\,|\,F]$ is called pure double circulant [4]. The connection between quasi-cyclic and pure double circulant is established by theorem 1.3 from [1]. The code C generated by $G=[P\,|\,Q]$ can also be generated by $G=[I\,|\,F]$, where $F$ is a circulant, iff $gcd(p(x),x^k-1)=1$. In such case the connection is $q(x)=p(x)f(x)\text{mod}(x^k-1)$.

The theorem 1.1 from [1] established that $rank[P\,|\,Q]=k-\deg(\gcd(p(x),q(x),x^k-1))$ in other words to avoid zero-weight codeword must satisfy $gcd(p(x),q(x),x^k-1)=1$.

The possible weight enumerators defined in [10] are

$$W_{70,1}=1+2\beta y^{12}+(11730-2\beta-128\lambda)y^{14}$$
$$+(150535-22\beta+896\gamma)y^{16}+\cdots$$

and

$$W_{70,2}=1+2\beta y^{12}+(9682-2\beta)y^{14}$$
$$+(173063-22\beta)y^{16}+\cdots$$

The singly even self-dual [70,35,12] code with parameters $\beta=416,\ \gamma=1$ was found in [10].

The singly even self-dual [70,35,12] codes with parameters γ = 0 and β=1012, 460,414, 368, 322, 276, 230, 184, and 138 are known from [12].

The singly even self-dual [70,35,12] codes with parameters β=230, 240, 250, 260, 270, 280, 290, 300, 310, 320, 330, 340, 350, 360, 370, 380, 390, 400, 410, 420, 430, 440, 450, 460, 470, 480, 490, 500, 510, 520, 530, 540 are known from [13].

The singly even self-dual [70,35,12] codes:

with weight enumerator $W_{70,1}$ and parameters: γ = 0, β = 112, 134, 156, 178, 200, 222, 244, 266, 288, 310, 332, 354, 376, 398, 420, 442, 464, 486, 508, 530, 552, 574, 596, 618; γ = 11, β = 618, 640, 662, 684 and 706; γ = 22, β = 684, 750, 772, 794;

with $W_{70,1}$, γ = 0, β = 88, 110, 132, 154, 176, 198, 220, 242, 264, 286, 308, 330, 352,374, 396, 418, 440, 462, 484, 506, 528;

with weight enumerator $W_{70,2}$ for β = 204, 226, 226, 248, 270, 270, 292, 314, 314, 336, 358, 358, 380, 402, 402, 424, 446, 468, 490, 490, 512, 534, 534, 556, 578, 600, 622, 644, 666, 798, 842 are known from [11].

The singly even self-dual [70,35,12] codes with weight enumerator $W_{70,1}$ and parameters γ = 0 and β = 102, 136, 170, 204, 238, 272, 306, 340, 374, 408, 442, 476, 510, 544, 578, 612 are known from [8].

All of these codes are not quasi-cyclic. As it was stated in [14] by exhaustive search there are no pure double circulant construction for singly even [70,35,12] self-dual codes.

In this paper we are able to find $p(x), q(x)$ such that $gcd\left(p(x), x^k - 1\right) \neq 1$, $gcd\left(q(x), x^k - 1\right) \neq 1$ but $gcd\left(p(x), q(x), x^k - 1\right) = 1$. This polynomials are used for generation a valid [70,35,12] singly even self-dual codes.

## II. Code construction

Let us consider the generator matrix produced by two circulants: the forward and the inverse with the same first row. It is easy to see that the resulting code will be self-dual. The standard inner product between the first and second row in the first circulant will be: $a = x_0 x_k + x_1 x_0 + x_2 x_1 + \cdots + x_k x_{k-1}$ and in the inverse circulant the result will be $a_{inv} = x_0 x_1 + x_1 x_2 + \cdots + x_{k-1} x_k + x_k x_0$. The $a = a_{inv}$ and the resulting sum will be $0$. This is true for any possible shifts. So, the polynomial pair $P_1 = [p_0, p_1, p_2, \ldots, p_{K-1}]$ and $P_2 = [p_{K-1}, p_{K-2}, \ldots, p_0]$ could be used for convolutional self-dual code generation. Further we will point out only the first polynomial. The second one will be obtained by inverse the first.

Let us note, that $x^{35} - 1$ has two divisors: $x^3 - x^2 - 1$ and $x^3 - x - 1$. In such case if $gsd\left(P_1(x), x^{35} - 1\right) = x^3 - x^2 - 1$ then $gsd\left(P_2(x), x^{35} - 1\right) = x^3 - x - 1$ due to symmetry and $gcd\left(P_1(x), P_2(x), x^k - 1\right) = 1$. We provide the exhaustive search for polynomials that satisfy the given conditions.

## III. Main result

We have obtained several codes with minimal weight codeword $d = 12$. All of codes have weight enumerator $W_{70,1}$ and $\gamma = 0$. These codes are listed below.

TABLE I. Code construction

| **Beta** | **Parameters** | | |
|---|---|---|---|
| | ***P*** | ***K*** | ***Number of ones*** |
| 140 | 11111101101 | 11 | 9 |
| 350 | 111011000101 | 12 | 7 |
| 420 | 111010000111 | 12 | 7 |
| 140 | 1110100011001<br>1100000111101 | 13 | 7 |
| 280 | 1110010001101 | 13 | 7 |
| 350 | 1100111010001 | 13 | 7 |
| 420 | 1100011101001 | 13 | 7 |
| 140 | 1110111010011 | 13 | 7 |
| 280 | 1110110010111 | 13 | 9 |
| 350 | 1111011100101<br>1111011010101 | 13 | 9 |
| 140 | 11110001001001 | 14 | 7 |
| 280 | 11001110100001 | 14 | 7 |
| 350 | 11110010010001<br>11101011000001<br>11100101010001<br>11100010101001<br>11011010000101<br>11001011001001<br>10110101010001 | 14 | 7 |
| 420 | 11010010110001<br>10110111000001 | 14 | 7 |
| 280 | 11110101011001<br>11101111100001 | 14 | 9 |
| 350 | 11110011100011<br>11100011101011<br>11100001111011<br>11011111000011 | 14 | 9 |
| 420 | 10111111001001 | 14 | 9 |
| 140 | 110011100001001<br>110011010001001 | 15 | 7 |
| 280 | 110101000010011 | 15 | 7 |
| 350 | 110101000001101<br>110100011100001<br>110010111000001<br>110000111100001<br>101110000001101<br>101100100100101<br>101100001101001 | 15 | 7 |
| 420 | 111010001000101<br>111000100100101<br>110100101000101 | 15 | 7 |
| 140 | 111100101100101<br>111010011100011<br>111000111101001<br>111000111010011 | 15 | 9 |
| 280 | 111001011000111<br>110111010011001 | 15 | 9 |
| 350 | 111011001011001<br>110111010001101<br>110110011010101<br>110101011011001 | 15 | 9 |
| 420 | 111110010101001<br>111100011010011<br>111100001100111<br>111001011101001<br>110110010011101<br>110101100110101 | 15 | 9 |
| 140 | 1110000000101011<br>1101000110010001<br>1101000001100101 | 16 | 7 |
| 280 | 1110010100010001<br>1101010010001001 | 16 | 7 |

| Beta | Parameters | | |
|---|---|---|---|
| | ***P*** | ***K*** | ***Number of ones*** |
| | 1101000000111001<br>1100011010000101<br>1100000101101001 | | |
| 350 | 1111100000010001<br>1111001000000101<br>1110001000100101<br>1110000010010011<br>1101100110000001<br>1101100001010001<br>1101010001000011<br>1100001101000011<br>1010101010010001<br>1010100101100001 | 16 | 7 |
| 420 | 1110001010000011 | 16 | 7 |

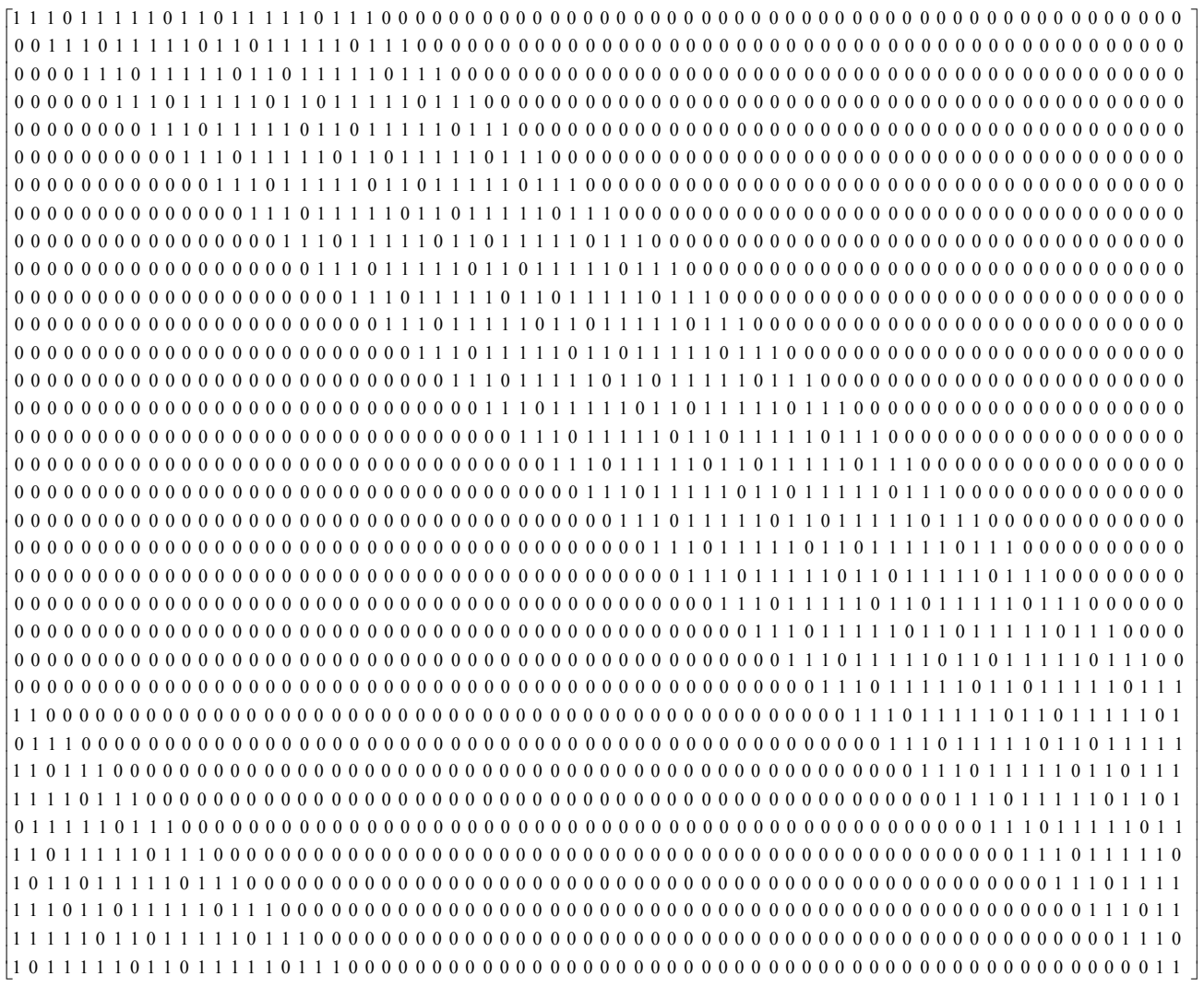

Fig. 1. The generator matrix G of self-dual tailbitting convolutional code for n=70, k=35, $P_1=(11111101101)_2$, $P_2=(10110111111)_2$ (inverse) with $\beta=140, \gamma=0$

The mixed polynomial string (non-zero part of the first row) is q=[1,1,1,0,1,1,1,1,1,0,1,1,0,1,1,1,1,1,0,1,1,1]. The odd positions are occupied by $P_1=[11111101101]$ and the even positions are occupied by $P_2=[10110111111]$.

## IV. Conclusion

We have obtained quasi-cyclic codes with m=2 with weight enumerator $W_{70,1}$ $\gamma=0$ and $\beta=140,280,350,420$. The codes with $\beta=140$ are the first known codes with these parameters.

## References

[1] Solomon G., van Tilborg H.C.A., “A connection between block and convolutional codes,” SIAM J. Applied Math., V37, N2, pp. 358-369, 10/79.

[2] Solomon G., “Convolutional encoding of self-dual codes,” Jet Propulsion Laboratory, Pasadena, CA, TDA Progress Report 42-116, pp. 110-113, 2/94.

[3] Solomon G., Jin Y., “Convolutional encoding of selfdual block codes (II),” Jet Propulsion Laboratory, Pasadena, CA, TDA Progress Report 42-118, pp. 22-25, 8/94.

[4] W.C. Huffman and V. Pless, Fundamentals of Error Correcting Codes, Cambridge University Press, 2003.

[5] J. H. Conway, N. J. A. Sloane, A new upper bound on the minimal distance of self-dual codes, IEEE Trans. Inform. Theory, 36(6), 1319-1333, 1990.

[6] E. M. Rains, Shadow Bounds for Self Dual Codes, IEEE Trans. Inf. Theory, 44(1), 134-139, 1998.

[7] Gabriele Nebe, Eric M. Rains, and Neil J. A. Sloane. 2006. Self-Dual Codes and Invariant Theory (Algorithms and Computation in Mathematics). Springer-Verlag New York, Inc., Secaucus, NJ, USA.

[8] Gürel, M., Yankov, N. (2016). Self-dual codes with an automorphism of order 17. Mathematical Communications, 21(1), 97-107. Retrieved from http://hrcak.srce.hr/157710

[9] W. Cary Huffman. 2005. On the classification and enumeration of self-dual codes. Finite Fields Appl. 11, 3 (August 2005), 451-490. DOI=http://dx.doi.org/10.1016/j.ffa.2005.05.012

[10] Masaaki Harada. The Existence of a Self-Dual [70, 35, 12] Code and Formally Self-Dual Codes. Finite Fields Appl. 3, 2 (April 1997), 131-139..

[11] N. Yankov, M.H. Lee, M. G¨urel and M. Ivanova, Self-dual codes with an automorphism of order 11, IEEE Trans. Inform. Theory 61 (2015), 1188–1193..

[12] Dontcheva, Radinka. "New Binary [70,35,12] Self-Dual and Binary [72,36,12] Self-Dual Doubly-Even Codes." Serdica Mathematical Journal 27.4 (2001): 287-302.

[13] Anne Desideri Bracco, Ann Marie Natividad, and Patrick Solé. 2008. On quintic quasi-cyclic codes. Discrete Appl. Math. 156, 18 (November 2008), 3362-3375..

[14] T. Aaron Gulliver and Masaaki Harada. 1998. Classification of Extremal Double Circulant Self-Dual Codesof Lengths 64 to 72. Des. Codes Cryptography 13, 3 (March 1998), 257-269.